\documentclass{elsart}
\usepackage{graphicx}
\usepackage{graphicx}
\usepackage{epsfig}
\usepackage{amsmath}
\usepackage{amssymb}
\usepackage[english]{babel}
\usepackage{textcomp}
\usepackage{lineno}
\usepackage{setspace}
\begin{document}
\begin{frontmatter}
\title{A $\mu$-TPC detector for the characterization of low energy neutron fields}

\author[irsn]{C. Golabek},
\author[lpsc]{J. Billard},
\author[irsn]{A. Allaoua},
\author[lpsc]{G. Bosson},
\author[lpsc]{O. Bourrion},
\author[lpsc]{C. Grignon},
\author[lpsc]{O. Guillaudin},
\author[irsn]{L. Lebreton},
\author[lpsc]{F. Mayet},
\author[irsn]{M. Petit},
\author[lpsc]{J.-P. Richer},
\author[lpsc]{D. Santos}

\address[irsn]{Laboratoire de M\'etrologie et de Dosim\'etrie des Neutrons, IRSN Cadarache, 13115 Saint-Paul-Lez-Durance, France}
\address[lpsc]{Laboratoire de Physique Subatomique et de Cosmologie, Universit\'e Joseph Fourier Grenoble 1, CNRS/IN2P3, Institut Polytechnique de Grenoble, 53 rue des Martyrs, 38026 Grenoble, France}

\begin{abstract}

The AMANDE facility produces monoenergetic neutron fields from 2~keV to 20~MeV for metrological purposes. To be considered as a reference facility, fluence and energy distributions of neutron fields have to be determined by primary measurement standards. For this purpose, a micro Time Projection Chamber is being developed to be dedicated to measure neutron fields with energy ranging from 8~keV up to 1~MeV. In this work we present simulations showing that such a detector, which allows the measurement of the ionization energy and the 3D reconstruction of the recoil nucleus, provides the determination of neutron energy and fluence of these neutron fields.

\end{abstract} 
\end{frontmatter}

\section{Introduction} 

For metrological purposes in ionizing radiation field, it is necessary to develop facilities producing, among other things, neutron fields. The AMANDE facility produces monoenergetic neutron fields from 2 keV up to 20 MeV~\cite{ama_gre}. To become a reference in neutron metrology and dosimetry, neutron fields have to be characterized in energy and fluence. Up to now, no detection system allows accurate measurement of the energy distribution of such low energy neutron fields (below a few tens of keV). For that purpose, a new gaseous detector system ($\mu$TPC) is being developed within the framework of the MIMAC collaboration~\cite{mimac} devoted to directional search for Dark Matter~\cite{wimp}~\cite{julien}~\cite{billard3}. Indeed, this search strategy requires track reconstruction of recoil nuclei down to a few~keV, which can be achieved with low pressure gaseous detectors~\cite{Ahlen}. Hence, the energy must be measured precisely and the track of the recoil nucleus must be reconstructed in 3D. For the $\mu$TPC, this is achieved with a low pressure micropattern gaseous detector (a pixelized bulk micromegas~\cite{Gio06}~\cite{esther}) equipped with a self triggered electronics system capable of performing anode sampling at a 40 MHz frequency~\cite{Ric06}~\cite{Bou06} (50 MHz with the new version of the electronics). This experimental device is also adapted for neutron metrology and designed to become a primary reference measurement standard in neutron fields measurement. Different detection conditions (gas, pressure, voltages) of the $\mu$TPC will allow measurement of neutron fields in the energy range from 8~keV up to 1~MeV.\\
Sec.~\ref{metrology} of this paper will describe the metrological context of these studies whereas the AMANDE facility itself will be presented in Sec.~\ref{amande}. The $\mu$TPC detection principle is described in Sec.~\ref{detector}. Based on simulated data, the calculation methods and expected performances for neutron fluence and energy determination will be shown in Sec.~\ref{fluence} and~\ref{energy} respectively.  

\section{Metrological context} 
\label{metrology}

Accurate dose measurement is important for radiation safety (e.g. nuclear industry, healthcare domain) of the workers as well as for medical treatments~\cite{CIPR60}~\cite{CIPR103}. As neutron dose equivalent depends on conversion coefficients (depending themselves on neutron energy), the determination of the fluence as a function of neutron energy is of first importance~\cite{ICRU39}. The Institute for Radiological Protection and Nuclear Safety (IRSN) owns the French references for the neutron fluence, dose equivalents and kerma quantities in neutron metrology~\cite{ICRU13}. These reference levels have been obtained by the development of neutron facilities as recommended by the ISO standards 8529-1~\cite{iso}, allowing characterization of survey instruments and dosimeters. Laboratories in other countries have their own reference standards~\cite{PTB}~\cite{NPL}. \\
Within the framework of French references, several facilities have been developed at IRSN. $^{241}$AmBe and $^{252}$Cf neutron sources are used for routine calibration. To adapt the response of dosimeters to workplaces, the CANEL assembly has been developed to produce realistic neutron radiation fields~\cite{CANEL}. However, as dosimeters response depends strongly on neutron energy, an accelerator producing monoenergetic neutron fields is needed. The AMANDE facility (Accelerator for Metrology and Neutron Applications in External Dosimetry) has been developed for this purpose~\cite{ama_gre}. It allows the study of dosimeter response variation for energies between 2 keV and 20 MeV. To consider AMANDE as a reference facility in neutron metrology, characterization of the neutron fields with respect to energy and fluence is undertaken using primary measurement standards. For a primary measurement standard, the procedure measurement used to obtain the measurand (energy or fluence) must be unrelated to a measurement standard of the same kind~\cite{VIM}. For example, a primary measurement standard dedicated to characterizing neutron fields is therefore not calibrated using a neutron field. \\
For the determination of the neutron fluence reference, a long counter, a polyethylene cylinder surrounding an $^{3}$He proportional counter, is used as reference measurement standard~\cite{counter}. For neutron energy, the reference measurement standard is based on the time-of-flight method~\cite{TOF}, which is relevant only for a pulsed ion beam. For a continuous beam mode, nuclear recoil spectrometers are used: a liquid scintillator BC501A~\cite{BC501A} and SP2 spherical proportional counters~\cite{SP2}. Calibrated at the \textsl{Physikalisch Technische Bundesanstalt} (PTB), they are therefore considered as secondary measurement standards. The $\mu$TPC designed to become a primary measurement standard in energy and fluence uses innovative technologies, an important feature in a metrological context. Indeed to approach the true quantity value~\cite{VIM}, it is necessary to use different measurement methods and thereby reduce uncertainties due to the applied methods. Present detectors used in primary facilities measure the neutron fluence with a maximum uncertainty of 5\% and typical uncertainty for energy is between 2\% and 8\%, both depending on the detectors used and energy measured. 

\section{Amande facility}
\label{amande}
\subsection{Neutrons production}

\begin{table}
\begin{center}
\begin{tabular}{|c|c|c|c|c|}
\hline
Neutron energy & Nuclear & Fluence rate & $\Delta E_n/E_n$ & Target thickness   \\
(keV) & reaction & $(cm^{-2}.s^{-1}.\mu A^{-1})$ & ($\%$) & ($\mu$g.cm$^{-2}$)  \\
\hline
8 & $^{45}Sc(p,n)^{45}Ti$ & 0.6 & 20 & 20  \\
27 & & 0.5 & 6 & 20  \\
\hline
144$^{(1)}$ & & 110 & 14 & 120  \\
250$^{(2)}$ & $^{7}Li(p,n)^{7}Be$ & 60 & 6 & 120  \\
565 & & 250 & 2 & 120  \\
\hline
\end{tabular}
\end{center}
\caption{Monoenergetic neutron fields of AMANDE (below 1~MeV) with the corresponding nuclear reaction. The fluence rates are given at 1 m from the target as well as the energy width for a given target thickness. Energy and fluence rate are given at 0$^\circ$ relative to the beam direction. See~\cite{ama_gre} for more details. References (1) and (2) are used in Fig.~\ref{dep_ang}.}
\label{energy1}
\end{table}

\begin{figure}[h]
\centering
\includegraphics[scale=0.8]{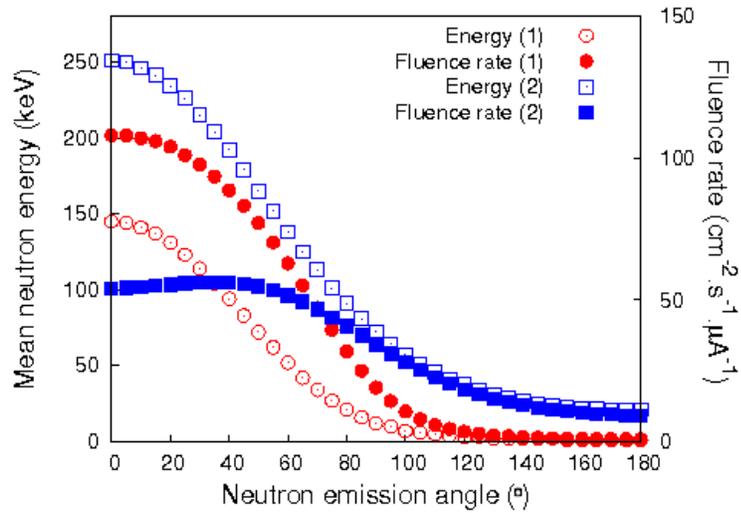} 
\caption{Variation of mean neutron energy and fluence rate (at 1 m from the target) as a function of neutron emission angle (relative to the ion beam direction). Simulation done with TARGET code~\cite{target} for the reaction $^7Li(p,n)^7Be$ producing neutron fields with mean energy 144 keV and 250 keV at 0$^\circ$ relative to the beam direction (configuration (1) and (2) of Tab.~\ref{energy1}).}
\label{dep_ang}
\end{figure}

The AMANDE accelerator is a 2 MV Tandetron system delivering proton or deuteron beams in the energy range from 100~keV up to 4~MeV, in DC or pulsed mode. Neutrons are produced by nuclear interactions of these ion beams with thin targets of scandium, lithium, deuterium or tritium. The targets are placed at the end of the beamline on a wobbling target holder allowing a circular scan of the beam spot onto the target, with an air cooling system.
Tab.~\ref{energy1} presents the monoenergetic neutron fields provided by AMANDE below 1~MeV; the corresponding nuclear reactions as well as the fluence rates and energy spread are shown for a given target thickness. The fluence rate is indicated as a rough guide at 1 m from the target at $0^\circ$ with respect to the incident beam direction, as calculated by the TARGET code~\cite{target} or from experimental data for scandium~\cite{Lam10}.\\
The Fig.~\ref{dep_ang} shows the variation of mean energy $E_n$ and fluence rate of the neutron fields as a function of the neutron emission angle for the two reactions (1) and (2) of Tab.~\ref{energy1}. Neutron fields are monoenergetic {\it i.e.} for each neutron emission angle, neutron energy is distributed within a unique peak centered at $E_n$ with a relatively small dispersion $\Delta E_n$, which depends on different factors: the energy spread of the ion beam (from 500~eV to 10~keV depending on beam energy), the energy broadening due to the beam energy loss inside the target as well as the beam energy stability. Another source of neutron energy broadening is the detection solid angle due to the dependence of both the neutron fluence and the energy with the emission angle. \\
Neutrons scattered in surrounding material contribute mostly to background neutrons and have to be limited. For that purpose, AMANDE has been built to minimize neutron scattering as decribed in Sec.~\ref{hall}. The amount of background neutrons produced by other reactions, such as neutrons produced by the reactions of beam on elements of the beam line or beam on the backing of the target, is comparable to natural background~\cite{gre_backing}. 

\subsection{Experimental hall} 
\label{hall}

In a metrological context, measurement accuracy is of first importance. In the AMANDE experimental hall, temperature and humidity match standard test conditions defined in the IEC standards~\cite{temp1}~\cite{temp2}. Neutron scattering is also minimized. Indeed, the $20 \times 20 \times 16$ $ m^3$ experimental hall has a floor grating placed 6 m above the ground over the entire hall surface with the exception of a 6 m-radius hole (cf. Fig.~\ref{hall}). In addition, the experimental hall is surrounded by metallic walls and ceiling. This design is allowed as AMANDE is situated at the center of a 50~m radius exclusion area. \\   
A fully automated transport system allows the positioning of detectors at any distance between 0.5 and 6~m and at an angle between $-150^\circ$ and $150^\circ$ from the neutron-producing target with respect to the incident beam direction. 

\begin{figure}[h]
\centering
\includegraphics[width=8.cm,keepaspectratio]{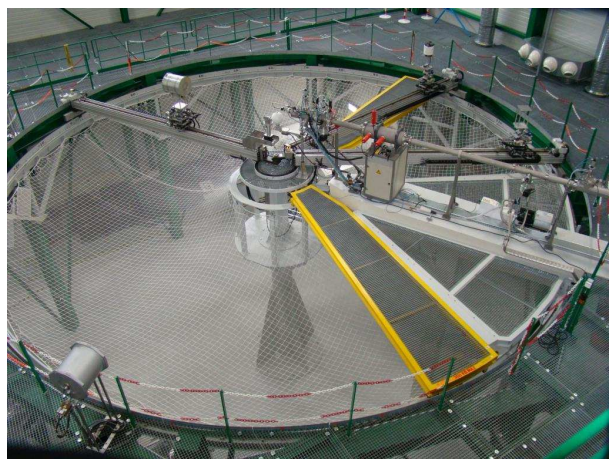} 
\caption{Experimental hall of AMANDE.}
\label{hall}
\end{figure}
\section{Principle and technical description of the $\mu$TPC}
\label{detector}

To measure the low energy neutron fields of AMANDE (below a few tens of keV), new dedicated detection devices are needed~\cite{amo}~\cite{amo_these}. A gaseous micro Time Projection Chamber ($\mu$TPC) is being developed within the framework of the MIMAC project which aims at 
detecting galactic
Dark Matter by directional detection \cite{mimac}~\cite{wimp}~\cite{julien}~\cite{billard3}. Changing the medium characteristics (gas mixture, pressure, voltages) and adapting the vessel of the detector to limit neutron scattering, this experimental device is adapted for 
neutron metrology based on the nuclear recoil telescope principle, {\it i.e.} the detection of a recoil nucleus
following the elastic scattering of the neutron in a target medium. The $\mu$TPC using a gaseous neutron converter provides a very low energy detection threshold, compared to, for example, a 5 MeV energy threshold for a CMOS proton recoil telescope with a 50~$\mu$m polyethylene converter~\cite{CMOS}~\cite{amo_these}. Using kinematic relations, the neutron energy 
$E_n$ is directly linked to the recoil energy $E_r$ and the recoil angle $\theta_r$ by : 
\begin{equation}
E_n = \frac{(m_n + m_A)^2}{4m_A m_n\times cos^2 \theta_r} \times E_r
\label{eq:en}
\end{equation}
where $m_A$ and $m_n$ are respectively the recoil nucleus mass and the neutron mass. The measured recoil energy $E_r$ has to be corrected by the ionization quenching factor (IQF) in order to get the total recoil energy (in a $C_4H_{10}$ $\mu$TPC operated at 50~mbar, IQF is about 99.6\% and 94\% for proton recoils of 144 keV and 8 keV respectively (SRIM simulations \cite{srim})). Measurements of IQF can be performed with a dedicated experimental facility at the \textsl{Laboratoire de Physique Subatomique et de Cosmologie} (LPSC)~\cite{IQF}. A good determination of $\theta_r$ suggest explicitly stating that the direction of incoming neutrons is known, which is the case at AMANDE facility. It follows that a simultaneous measurement of the energy
and angle of the recoil nucleus leads to the determination of the initial neutron energy, as long as the recoiling nuclear species is known.\\

\begin{figure}[h]
\centering
\includegraphics[scale=0.5]{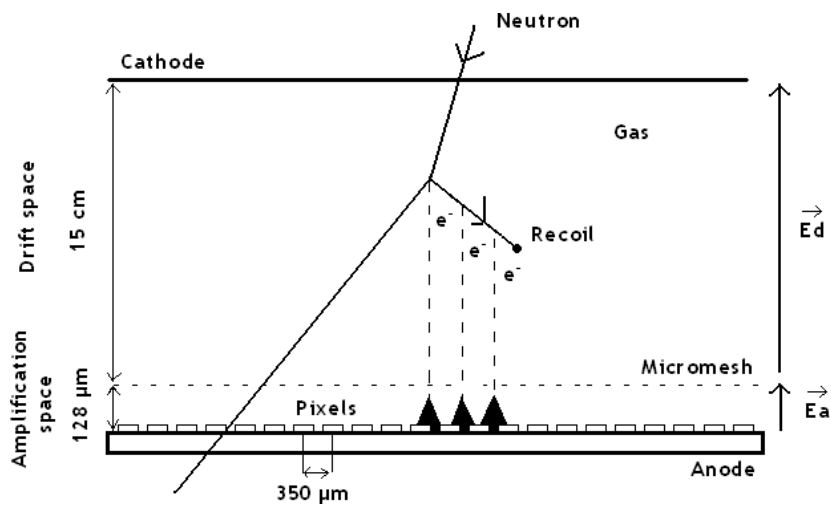} 
\caption{Sketch of the $\mu$TPC.}
\label{principe}
\end{figure} 
 
The primary electron-ion pairs produced by a nuclear recoil in the $\mu$TPC are detected by drifting the electrons 
to the grid of a bulk micromegas \cite{Gio06}
 and producing an avalanche in the amplification space with a very thin gap (128 or 256~$\mu$m, cf. Fig.~\ref{principe}). 
 The electrons move towards the grid in the drift space and are projected on the anode thus providing information on X and Y coordinates.
To access the X and Y dimensions with a 100 $\mu$m spatial resolution, a bulk micromegas with a 4 by 4 cm$^2$ active area (10 by 10 cm$^2$ in the new version), segmented in pixels with a pitch of 350 $\mu$m is used as 2D readout. A uniform distribution of the electrons has been supposed between pixels as an approximation of the barycenter of the 2D projection on the anode.
In order to reconstruct the third dimension Z of the recoil, self-triggered electronics have been developed. It allows performance of anode sampling at a frequency of 40 MHz (50 MHz in the new version of the electronics), producing slices of X and Y positions every 25 ns (20 ns in its new version, cf. Fig.~\ref{slices}).
This system includes a dedicated 16 channel ASIC \cite{Ric06} (64 in its new version) associated with a DAQ \cite{Bou06}.
The Z coordinate is then obtained
using the drift time, assuming that the electron drift velocity is
known. The drift velocity may be measured by collecting the primary electrons produced by alpha particles ($^{241}$Am source for example) crossing the chamber between the cathode and the anode or evaluated with the MAGBOLTZ sofware~\cite{magboltz}. Preliminary measurements of the drift velocity using alpha particles in different gases have been performed and fit well with MAGBOLTZ simulations~\cite{amo_these}. A 3D reconstruction algorithm has been developed and sucessfully tested on
simulation and experimental data. More details may be found in \cite{billard.track}~\cite{confTPC}. \\  
Several gas mixtures ({\it e.g.} $C_{4}H_{10}$, $C_{4}H_{10}$ + $CHF_{3}$ or $^{4}He$ + $C_{4}H_{10}$) may be used with the $\mu$TPC, with several pressures. The choice is mainly a question of
track length and energy threshold (note that the track length of a 144~keV proton in a $C_4H_{10}$ $\mu$TPC operated at 50~mbar is about 1~cm). A hydrogen target is more convenient than heavier nuclei: the neutron scattering cross section on hydrogen and helium vary from 19.5 barn to 4 barn and from 7 barn to 0.7 barn respectively in the energy range 8~keV~-~1~MeV~\cite{endf1}~\cite{endf2}. Moreover the high kinetic energy and long range of hydrogen in gas improve energy resolution and ease identification. 
In the case of multitarget mixtures, recoil identification may be performed thanks to the combined use of energy and track length information \cite{Discri}.

\section{Measurement of the neutron fluence}
\label{fluence}

The detection of recoil nuclei created in the $\mu$TPC allows us to calculate the angular neutron flux $d\varphi/dt (\theta, \phi)$ ($s^{-1}sr^{-1}$). The low pressure of the medium (around hundred mbar) and values of neutron scattering cross sections imply that the probability for a neutron to produce more than one recoil is about 10$^{-3}$. Then in the thin slice approximation, the recoil nuclei rate $dN_{int}/dt$ ($s^{-1}$) is given by:

\begin{equation}
\frac{dN_{int}}{dt} =  \sigma(E_n) \frac{\rho N_A n}{M_{mol}}  \int_{V_D} k(E_n,z) \frac{d\varphi}{dt}(\theta, \phi) \sin{\theta}drd\theta\d\phi
\label{eq:def}
\end{equation}

where $\sigma(E_n)$ is the neutron scattering cross-section, $N_A$ is the Avogadro number, $\rho$ and $M_{mol}$ the density and atomic weight of the medium, $n$ the number of relevant atoms in the molecule of the medium ({\it e.g.} $n$=10 using $C_{4}H_{10}$ and considering proton recoils) and $k(E_n,z)$ the correction factor~\cite{gum}. $k(E_n,z)$ is related to the loss of events due to detection (track length, threshold) or data analysis (selection cuts, fiducial volume), then depends on nature of gas, pressure and neutron energy range. The value of $k(E_n,z)$ cannot be determined by measuring a characterized neutron field as the $\mu$TPC is intended to become a primary detector standard (cf. Sec.~\ref{metrology}); it shall be evaluated by a full Monte Carlo simulation (Geant 4, MCNPX) including DAQ and thresholds. Parameters used in the simulation will be determined by experimental measurements using an ion source specially developed to measure the quenching factor~\cite{IQF} and electron recombination (depending on z). The integration is performed
on the detection volume $V_D$, calculating beforehand the mean value $\bar{k}(E_n)$. If $d\varphi/dt (\theta, \phi)$ can be considered as constant throughout the detection volume, providing the variation of the fluence as a function of angle is small in the solid angle of the $\mu$TPC (cf. Fig.~\ref{dep_ang}, assumption validated with simulation (cf. below)), the angular neutron flux is directly linked to the recoil rate via the geometry efficiency $\epsilon_{geo}$ defined as:

\begin{equation}
\epsilon_{geo} = \int_{V_D} \sin{\theta}drd\theta\d\phi
\end{equation}

\begin{figure}[h]
\begin{center}
\includegraphics[scale=0.6]{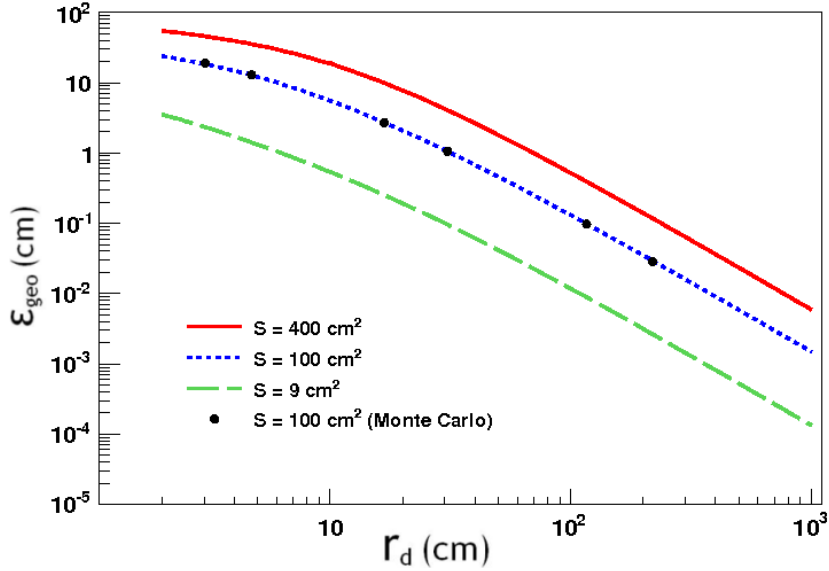} 
\caption{$\epsilon_{geo}$ as a function of $r_d$ (source-detector 
distance) for a 15 cm long detector with 3 different anode sizes (400, 100 and 9 cm$^2$). Lines present results from numerical integration while
points present results from Monte Carlo simulation.} 
\label{fig:EpsilonGeoVsX1bis}
\end{center}
\end{figure}

The value of $\epsilon_{geo}$ is evaluated either by numerical integration or by Monte Carlo simulation. It depends only on the detector configuration, i.e. both the detector geometry and the source-detector 
distance ($r_d$). Fig.~\ref{fig:EpsilonGeoVsX1bis} presents  $\epsilon_{geo}$ as a function of $r_d$ for a 15~cm long detector with 3 different anode sizes (400, 100 and 9~cm$^2$). It may be noted that at large distance, $\epsilon_{geo}$ scales as $1/r_d^2$ as expected. \\
Finally, the neutron fluence rate $d\Phi/dt(r\verb+|+\theta_d)$ ($cm^{-2}.s^{-1}$) is calculated via the expression~(\ref{eq:def2}), at a given position $r$ from the neutron producing target and for a given polar angle $\theta_d$ with respect to the incident beam direction. $d\Phi/dt (r\verb+|+\theta_d)$ is therefore the neutron fluence rate averaged over the whole solid angle of the $\mu$TPC. 

\begin{equation}
\frac{d\Phi}{dt}(r\verb+|+\theta_d)  =  \frac{1}{r^2} \times \frac{1}{\bar{k}(E_n) \epsilon_{geo}} \times \frac{M_{mol}}{\rho  N_A {\sigma(E_n)}n} \times \frac{dN_{int}}{dt}    
\label{eq:def2}
\end{equation}

Note that this holds true if $\sigma(E_n)$ may be considered as a constant on the detection volume. Using a 144~keV neutron source (for which the absolute energy dispersion is the larger (cf. Tab.~\ref{energy1})) and placing the $\mu$TPC at 30~cm from the source, the neutron energy dispersion is around 15\% (cf. Tab.~\ref{resolution}). This leads to a variation of the neutron scattering cross-section of 3\%. A MCNPX simulation (MCNPX-2.6f version) has been performed using a 144~keV neutron source (TARGET code) realistic for AMANDE. Neutron energy and fluence of the neutron field depend therefore on the $\theta$ angle according to Fig.~\ref{dep_ang}. In our simulation the $\mu$TPC was placed at 10 cm from the neutron source with an anode of 100 cm$^2$. Even in this unfavorable configuration (practically the $\mu$TPC will be placed at least at 30 cm from the neutron source), the neutron fluence has been determined with an error less than 1\% which validates the expression~(\ref{eq:def2}) and the associated assumptions: in the whole solid angle of the detector, the neutron scattering cross section can be considered as constant and the calculated averaged neutron fluence gives an accurate estimation of the neutron fluence at the angle $\theta_d$. \\
To control the atomic density $\rho$ of the medium and to remove contaminants, a gas controller allowing to manage the pressure and the renewal of the gas in the chamber is being built. For the neutron scattering cross sections $\sigma(E_n)$, nuclear data bases provide values for (n,p) reactions  with uncertainty less than 1\%~\cite{sigma1}~\cite{sigma2} in the considered energy range.\\
The finite size of the detection volume implies geometrical effects, i.e. the recoils may not be fully contained in the detection volume. It follows that their kinetic energy is not correctly measured and these events should not be taken into account, thus reducing the efficiency. A high gas pressure would then be suitable for limiting geometrical effects. Nevertheless the determination of the angle depends on the track reconstruction {\it i.e.} the number of sampling slices of the recoil tracks (cf.~Sec.~\ref{detector}); a low gas pressure leads then to long tracks. Consequently a compromise has to be reached. Experimentally a $C_4H_{10}$ $\mu$TPC operated with pressures between 50~mbar and 200 mbar has been successfully tested. For neutron fields produced by AMANDE (cf. Tab~\ref{energy1}), to measure 8, 24 and 144~keV neutron fields, a $C_4H_{10}$ $\mu$TPC operated at 50~mbar may be used (Note that below 50 keV, $CHF_3$ has to be added to $C_4H_{10}$, c.f.~\ref{energy}). For 250~keV and 565~keV neutron fields, 100~mbar and 200~mbar are respectively relevant. In these configurations the outgoing recoils represent less than 20\% of the total recoils. 
Assuming a correction factor $k$=0.8 (i.e. based only on the geometrical loss of 20\% of the recoils), the ratio of measured proton recoils with respect to incident neutrons is about $10^{-3}$ in these configurations. Taking into account neutron fields provided by AMANDE and with a $\mu$TPC placed at 30~cm from the target, this leads to an irradiation time of less than 10~min to have a statistical uncertainty of 1\% for 144, 250 and 565~keV neutron fields with $I_{beam}$=$1\mu A$, and less than 25~min for 8, 27~keV neutron fields with $I_{beam}$=$20\mu A$. This uncertainty corresponds to 10000 detected recoils including only geometrical efficiency. This calculation gives roughly an estimation of the time needed to characterize the neutron fields produced at AMANDE facility.

\section{Measurement of the neutron energy}
\label{energy}
\begin{figure}[t]
\begin{center}
\includegraphics{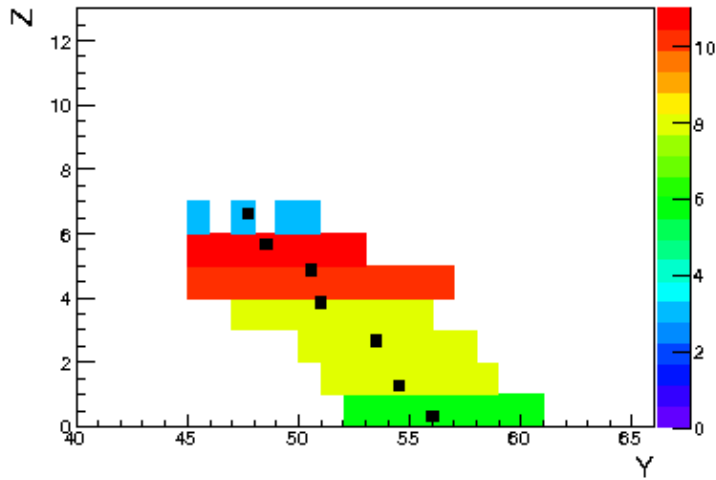}
\caption{Reconstructed track of a simulated 100 keV proton recoil with $\theta=\phi=45^\circ$ in 50 mbar $C_{4}H_{10}$ projected on the YZ plane.}
\label{slices}
\end{center}
\end{figure}

\begin{figure}[t]
\begin{center}
\hspace*{-2cm}
\includegraphics[scale=0.8]{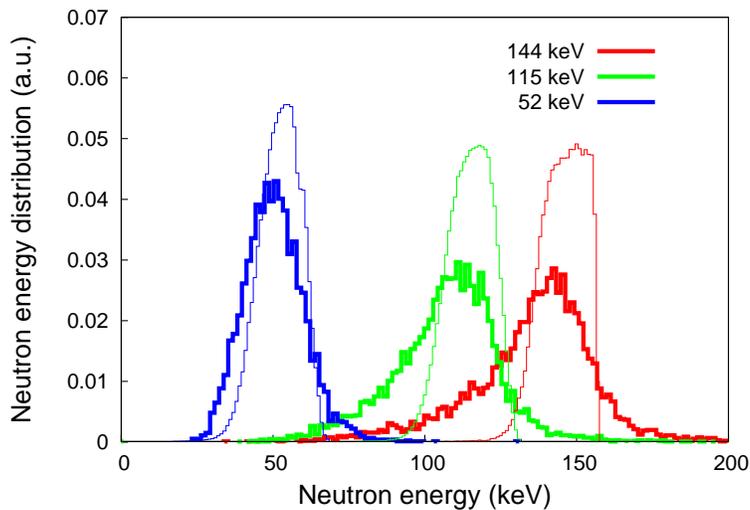}
\caption{Energy distributions of neutrons going through the detection volume of the $\mu$TPC for three neutron mean energies, as produced by the neutron source (thin line distributions) ; a neutron field of 144 keV (configuration 1 of Tab~\ref{energy1}) at 0{$^\circ$} leads to a mean neutron energy of 115 keV at 30{$^\circ$} and 52 keV at 60{$^\circ$}. Thick lines distributions show the corresponding reconstructed neutron energy distributions. For the distribution of 52 keV, there is no tail at low energy due to the threshold of the reconstruction at 10 keV.}
\label{144}
\end{center}
\end{figure}

\begin{figure}[t]
\begin{center}
\includegraphics[scale=0.8]{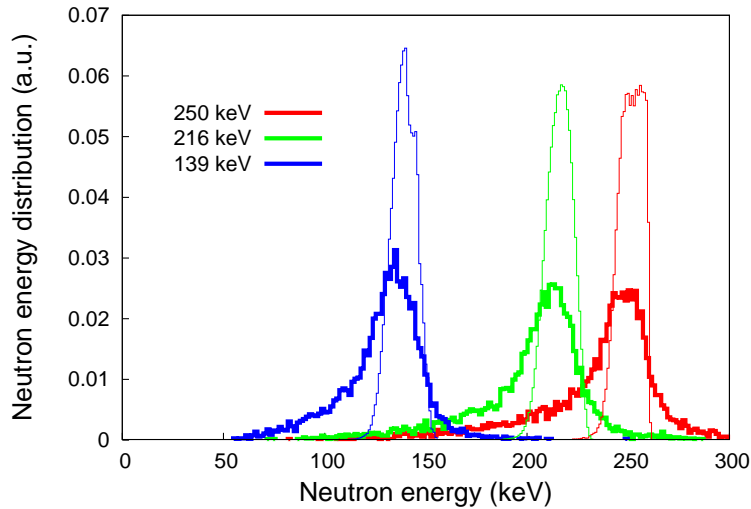}
\caption{Same distributions as Fig.~\ref{144} but a neutron field of 250 keV at 0{$^\circ$} (configuration 2 of Tab~\ref{energy1}).}
\label{250}
\end{center}
\end{figure}

As stated above (cf. Sec~\ref{detector}), the simultaneous measurement of the scattering angle and the energy of the recoil nucleus leads to the determination of the neutron incident energy. In the following, a comprehensive data simulation has been performed to evaluate the expected performances on a monoenergetic neutron field, such as the one delivered at the AMANDE facility.  
The neutron fields (configurations (1) and (2) of Tab.~\ref{energy1}) are simulated with the TARGET code coupled to a MCNPX code (cf. Fig.~\ref{144} and~\ref{250}: thin line distributions) and used to produce induced recoils, assuming an isotropic elastic scattering in the center of mass frame~\cite{amo_these}.
Once the energy $E_r$ and the angle $\theta_r$ of the recoil are chosen, an event is randomly generated with track simulation software using
SRIM (simulated tracks), MAGBOLTZ \cite{magboltz} (electron drift velocity and dispersions) combined
with the detector response simulation (MIMAC DAQ). Note that the longitudinal electron diffusion, not taken into account in this simulation, will be included in the next step of the simulation studies.\\
In each configuration,  10000 simulated recoils are generated. In order to obtain for each recoil both the 3D track and the recoil energy, the analysis software is then used. It is worth noting that the ionization quenching factor is estimated from SRIM simulations, as described is Sec.~\ref{detector}. 
 The following selection is applied :
 \begin{itemize}
 \item Geometrical cuts :  the track must be enclosed in the detection volume, otherwise the energy measurement would be wrong.
 \item Track length : the track length must have at least 3 time slices in order to have a correct estimation of the recoiling angle. Fig.~\ref{slices} presents a reconstructed track of a simulated 100 keV proton recoil with $\theta=\phi=45^\circ$ in 50 mbar $C_{4}H_{10}$. In this case, the track has 7 slices.
 \end{itemize} 

Neutron energy is calculated with expression~(\ref{eq:en}), hereafter referred to as the {\it reconstructed neutron energy}. The results are shown on Fig.~\ref{144} and~\ref{250} (thick line distributions); neutron fields of configurations (1) of Tab.~\ref{energy1} (and (2)) have been considered at three different positions of the detector: 0{$^\circ$}, 30{$^\circ$} and 60{$^\circ$} with respect to incident beam direction corresponding to three different neutron fields with a mean energy of 144, 115, and 52~keV (and 250, 216 and 139~keV) respectively. In each case, the mean energy of the distribution is well reconstructed, with an error smaller than 7\%. 
In order to quantify the reconstruction ability of the detector, the full width at half maximum of neutron energy distributions from the neutron source and after reconstruction are compared in Tab.~\ref{resolution}. When the neutron mean energy increases from 50 to 250 keV, the neutron energy resolution (FWHM) decreases from 40\% to 10\%. Part of the dispersion of the reconstructed neutron energy is due to the dispersion of the neutron field itself (from 6.8\% to 32.7\%). The other component of this dispersion comes from the angular and energy dispersion due to reconstruction and detection methods (estimated at about 10\% simulating a Dirac energy distribution of 144~keV). 
 
\begin{table}
\begin{center}
\begin{tabular}{|c|c|c|c|c|c|c|c|}
\hline
Neutron energy E$_{mean}$ (keV) & 250 & 216 & 139 & 144 & 115 & 52 & 8 \\
\hline
Detector position & 0{$^\circ$} & 30{$^\circ$} & 60{$^\circ$} & 0{$^\circ$} & 30{$^\circ$} & 60{$^\circ$} & 0{$^\circ$} \\
\hline
FWHM/E$_{mean}$ source (\%) & 6.8 & 7.4 & 10.8 & 14.6 & 16.5 & 32.7  & 0.4 \\
\hline
FWHM/E$_{mean}$ reconstructed (\%) & 10.4 & 12.5 & 21.6 & 18.6 & 21.7 & 38.5 & 1 \\
\hline
\end{tabular}
\end{center}
\caption{For each neutron distribution (from source and after reconstruction) of Fig.~\ref{144},~\ref{250} and ~\ref{8kev}, the full width at half maximum (FWHM) in \% is given.}
\label{resolution}
\end{table}

\begin{figure}[t]
\begin{center}
\includegraphics[scale=0.8]{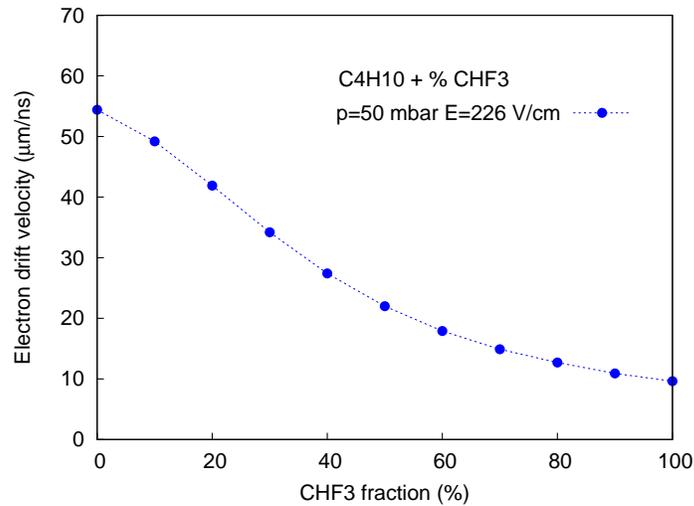}
\caption{Electron drift velocity as a function of $CHF_{3}$ gas content in $C_4H_{10}$ + $CHF_{3}$ gas mixtures for P = 50 mbar and E = 226 V/cm.}
\label{Vd}
\end{center}
\end{figure}

\begin{figure}[t]
\begin{center}
\includegraphics[scale=0.8]{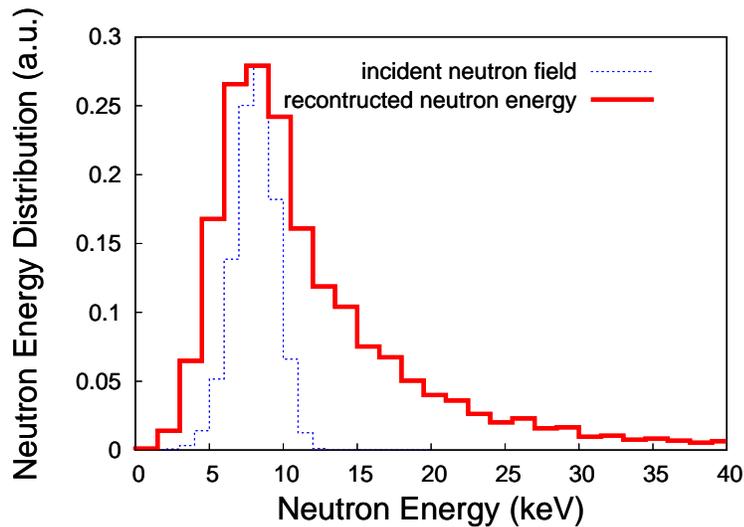}
\caption{Energy distribution of neutrons going through the detection volume of the $\mu$TPC placed at 0{$^\circ$} relative to the beam direction, as produced by the neutron source of mean energy 8 keV delivered by the AMANDE facility. Red line distribution shows the corresponding reconstructed neutron energy distribution.}
\label{8kev}
\end{center}
\end{figure}

Pure $C_4H_{10}$ can only be used for high energies (50 keV - 1 MeV). For low energies or to improve the energy reconstruction, a mixed gas ($C_4H_{10}$ + X\% $CHF_{3}$) can be used to reduce the drift velocity. The figure~\ref{Vd} shows that the drift velocity of electrons is comprised between 54 $\mu$m/ns and 10 $\mu$m/ns for 0\% and 100\% of $CHF_{3}$ respectively at 50 mbar. As example, the drift velocity is 18,3 $\mu$m/ns with a $C_4H_{10}$ + 50\% $CHF_{3}$ $\mu$TPC operated at 50 mbar. In this configuration, the track length of 8 keV protons is roughly 1.6 mm leading to 4 time slices. In order to show the ability of the $\mu$TPC using this gas mixture, the 8 keV neutron field produced by AMANDE (cf. table~\ref{energy1}) is simulated with the TARGET code coupled to a MCNPX code (cf. dotted blue line in figure~\ref{8kev}); the $\mu$TPC is placed at 0{$^\circ$} relative to the beam direction. The reconstruction of the energy distribution is performed in the same way than previously for the 144 and 255 keV neutron fields (cf. red distribution in figure~\ref{8kev}). Assymetry and dispersion of the reconstructed energy distribution is due to reconstruction and detection methods (cf. value of FWHM on table~\ref{resolution}). The maximum of the energy distribution is nevertheless preserved. To improve the track reconstruction, a maximum likelihood analysis~\cite{billard.track} will be applied to the raw data to determine the recoil track, in order to take into account also the longitudinal 
diffusion of the primary electrons collected to the anode. Moreover, a possible reduction of the pressure up to 30 mbar can be envisaged increasing the track length of protons.

\section{Conclusion} 

A micromesh gaseous detector $\mu$TPC has been developed within the MIMAC project. Using simulations, this paper describes and validates the methods to determine both the fluence rate and the neutron energy of neutron fields. A comprehensive data simulation showed the expected energy spread after reconstruction of neutron fields produced by the AMANDE facility between 50 and 250 keV using pure $C_4H_{10}$. For lower energy, gas mixture composed of $C_4H_{10}$ + X\% $CHF_{3}$ should be used. Experimental tests will be soon published. As a conclusion, this study shows that the $\mu$TPC opens the possibility to characterize low energy neutron fields like the ones produced by AMANDE. 
 
\section{Acknowledgements} 

This work was supported by the french National Laboratory of Metrology (LNE) and by the National Research Agency through ANR-07-BLANC-0255-03.

\end{document}